\documentstyle[nanosymp,epsf]{article}
\begin{document}

\title{\centering{Spontaneous Magnetization of a Two-Dimensional \\Electron Gas}}

\author{V.~V'yurkov  and A.~Vetrov}
\affil{Institute of Physics and Technology, Russian
Academy of Sciences,\\
Nakhimovsky prosp. 34, Moscow, 117218, Russia\\
Phone: (095)3324918, Fax: (095)1293141
E-mail: vyurkov@ftian.oivta.ru}

\beginabstract
\it{
Spontaneous magnetization of a two-dimensional electron gas (2DEG) is discussed. 
It takes place for sufficiently high electron density 
(i.e. for a quantum fluid state) with $r_s<10$ which is quite beyond the condition of
 Wigner crystallization $(r_s>37)$ obtained by Tanatar and Ceperley. The effect 
essentially depends upon screening and disorder . The energy interval under 
the Fermi level where a spin polarization occurs as a function of electron
 density is computed. The spontaneous magnetic moment and a decrease of
 electron density of states (pseudo-gap) at the Fermi level are found.}
\endabstract

The investigation of a 2DEG are still in focus of modern physics. 
Here we discuss a possibility of spontaneous magnetization of a 2DEG 
due to exchange interaction.
The entirely magnetized state of a 2DEG is a Wigner crystal. Tanatar
 and Ceperley in their earlier paper [1] computed the condition for a
 two-dimensional electron fluid to cross over to the Wigner crystal 
state [2]  at zero temperature. They obtained that the Wigner crystal  
melts  at $r_s=37\pm5$ where $r_s$ is the Brueckner parameter  
$r_s=\alpha_W/\alpha_0,\alpha_0={\hbar}^2\kappa/me^2$  
 being an effective Bohr radius (here $\kappa$ is a permittivity) and $a_W$ is the 
Wigner-Seitz radius connected with an electron sheet density $\sigma$ by 
a relation $\alpha_W=1/\sqrt{\pi\sigma}$ . Recently Yoon, et. al. [3]
 have observed some phase transition at   which they attributed to the 
melting of a Wigner crystal. 

Here we deal with a comparatively high electron density $(r_s$ less than $\approx37)$ 
corresponding to a homogeneous quantum fluid. The authors of  Ref.[1] 
investigated only the transition between ferromagnetic Wigner crystal 
and unpolarized  quantum fluid state. Thus they have not put into account 
a partially polarized state which is just the goal of the present consideration. 

Firstly, it was pointed out to an exchange interaction as a reason of spontaneous
 spin polarization in quantum wires by Wang and Berggren [4] with the aim to explain
 anomalies in quantum wire conductance. Their approach was based on Kohn-Sham local
 density description of the exchange interaction. Further, this approach was
 developed in Ref. [5,6]. However, in spite of simplicity, this method is 
 insufficient to reveal subtle details of spin configuration of electron system, 
for instance, to derive density of states at the Fermi level essential for any 
transport phenomena. 

In our model of a quantum fluid a great number of electrons are confined in 
a square region. At the very beginning, the problem of two-particle interaction 
is preliminary considered with Hartree-Fock equations similar to the case of a 
quantum wire [7-9]. It allows to introduce an exchange interaction in a many-particle
 Hamiltonian.

The exchange energy of two electrons is assumed to be small compared with kinetic one.
 Therefore, the Hartree-Fock (HF) approach is employed to describe exchange interaction.  
In this approach the energy of spin-less Coulomb interaction for two electrons 
in the region is as follows

\begin{equation}
V_c(\vec k_1,\vec k_2)={e^2\over{\kappa}}\int\!\int{\vert\psi_1(\vec k_1,\vec r_1)\vert^2
\vert\psi_2(\vec k_2,\vec r_2)\vert^2\over\vert\vec r_1-\vec r_2\vert}d\vec r_1d\vec r_2 
\end{equation}		

The  exchange energy magnitude is

\begin{equation}
\!\!\!\!V_{ex}\!(\vec k_1,\vec k_2)=\pm\left({e^2\over{2\kappa}}\int\!\!\!\!\!\int{\psi_1^*(\vec k_1,\vec r_1)\psi_1(\vec k_1,\vec r_2)
\psi_2^*(\vec k_2,\vec r_2)\psi_2(\vec k_2,\vec r_1)\over\vert\vec r_1-\vec r_2\vert}d\vec r_1d\vec r_2+c.c. \right) 
\end{equation}	

where $\psi_1, \psi_2$ are wave functions of the first and second electron normalized 
per one electron in the area. The sign of exchange energy depends upon spin state of the electron pair.
The unperturbed one-particle wave function is

\begin{equation}
\psi(\vec k,\vec r)={{1\over L}e^{i\vec k\vec r}}
\end{equation}    		

 where $L$ is a normalization length, $\vec r=(x,y)$ is a position.
                                                                   
Two electrons with sufficiently small momentum discrepancy  ${\hbar}\Delta 
k={\hbar}\vert \vec k_1-\vec k_2\vert$ so that

\begin{equation}
h\Delta k<h/\lambda
\end{equation} 
                                                             		
where $\lambda$ is an effective screening length $(\lambda<<L)$ possess an 
exchange energy (2) almost as large as Coulomb one (1). In particular,
 screening might be caused by mirror charges in  adjacent electrodes.
 In this case for rough estimations $\lambda$ could be evaluated as a distance
 from a 2DEG to the nearest electrode. 

In further calculations with equ.s (1) and (2) the Coulomb potential 
  was cut off for distances x larger than effective screening length $\lambda$.
 
For greater momentum mismatch than that given by the inequality (4) 
the exchange integrals (2) involve fast oscillating functions and tend to zero. 
Worth mentioning that a similar effect could be caused by disorder or 
scattering which break a wave function phase and thus diminish an overlap integral (2). 
For approximate evaluations in the case a screening length could be replaced by a 
coherence length $\lambda_\phi$. 

A sign of exchange energy in the exp.(2) depends upon spin configuration. If electrons
 have an antisymmetric spin configuration (total spin equals unity) then their space 
wave function is symmetric and the sign of exchange energy is positive, i.e. the same
 as that of a Coulomb energy.  Otherwise, when a total spin equals zero, the exchange 
energy is negative and reduces total energy of electron system. For the sake of 
simplicity in further calculations we suppose the exchange energy to be equal to the
 Coulomb one (1) when the condition (4) is true. Otherwise, it is supposed to equal zero.

The above model of exchange interaction was employed to solve a many-electron problem.
 It was assumed that electrons with one spin orientation occupy the states in the
 momentum  $\vec k-space$ up to $|\vec k|\rm=k^\uparrow$. At the same moment,
 electrons with the opposite spin
 orientation occupy the states up to $|\vec k|\rm=k^\downarrow$.  To characterize 
this spin configuration
 we introduce an energy $\delta\epsilon=({\hbar}^2/2m)|(k^\uparrow)^2-(k^\downarrow)^2|$.
 In absence of spin polarization (magnetization) $k^\uparrow=k^\downarrow$ 
and $\delta\epsilon=0$. The value of $\delta\epsilon$ corresponds to the energy 
interval under the Fermi level where a spin polarization arises. 

The exchange energy is calculated from the equation (2) and the supposition (4).
 The latter defines a circle of nearby electrons in $\vec k-space$ where an exchange
 interaction is strong. The sign of this interaction much matters, for example, 
it results in full compensation of exchange energy for deep states in $\vec k-space$. 
Moreover, the exchang energy of any electron in unpolarized fluid is fully compensated
 too (equals zero). It readily contradicts with the Kohn-Sham description.

The total energy of the electron system reduced per one electron $E_{tot}$ as a function 
of a parameter $\delta=\delta\epsilon/\epsilon_F$ ($\epsilon_F$ is the Fermi energy) was estimated.
 Evidently, the value of $E_{tot}$ includes the kinetic, Coulomb and exchange energy.
 In Rydberg units $Ry={\hbar}^2/2ma_0^2$  $E_{tot}$ reads

\begin{equation}
E_{tot}(\delta)\approx(4\pi^2/r_s^2)Ry(1+\delta^2/4-1/2\pi^3(L/a_0)RyE_{ex}f(\delta))+E_c
\end{equation} 

where

\begin{equation}
\!\!\!\!\!\!\!\!f(\delta)\!=\!\!\!\!\!\!\int\limits_{\sqrt{1-{\delta\over2}}}^{\sqrt{1+{\delta\over2}}}\!\!\int
\limits_{\sqrt{1-{\delta\over2}}}^{\sqrt{1+{\delta\over2}}}\!\!\!\!\!t_1t_2dt_1dt_2\!\!\int\limits_0^{2\pi}
\!\!\int\limits_0^{2\pi}\!\!\theta\!\left(1-2\pi{\lambda\over{\lambda_F}}\sqrt{t_1^2+t_2^2-2t_1t_2cos(
\alpha_1-\alpha_2)}\right)\!d\alpha_1d\alpha_2
\end{equation}

is a dimensionless function arising from the exp. (2) and inequality (4), $E_{ex}$ is a value 
of the integral (2) in the limit $|\vec {k_1}-\vec {k_2}|\rm\to 0$, the Coulomb energy $E_c$ does not
 depend upon spin configuration, i.e. on the parameter $\delta$.

If the condition $\lambda<<L$ is true the value of $E_ex$ can be estimated $E_{ex}\approx6.1(L/\lambda)$.
 The expression (5) may be thus written in more explicit form

\begin{equation}
E_{tot}(\delta)\approx(4\pi^2/r_s^2)Ry(1+\delta^2/4-0.01(\lambda/a_0)f(\delta))+E_c
\end{equation} 

The integral (6) was calculated numerically by the Monte Carlo method for typical values of screening 
length $\lambda$. It was found out that the ground state (i.e. for zero sample temperature T=0) corresponding 
to the minimum of the total energy (7) can be that of spin-polarized electrons near the Fermi level 
when $r_s>2$. The dependence of the parameter $\delta=\delta\epsilon/\epsilon_F$  characterizing 
spin-polarisation degree against the Bruekner parameter $r_s$ is depicted in the Fig.1. In contrast 
to 1DEG (Ref. [7-9]) a smooth transition to spin-polarized state has been obtained for 2DEG.

\begin{figure}[t]
\leavevmode
\centering{\epsfbox{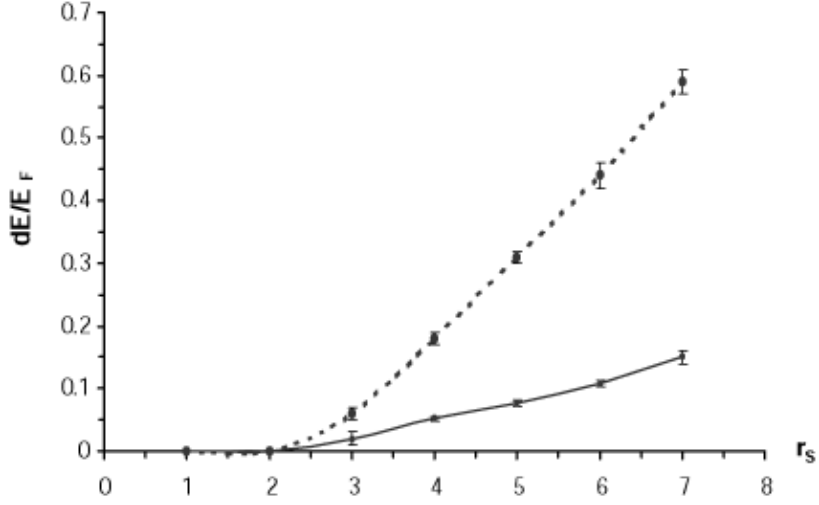}}
\caption{The parameter $\delta=\delta\epsilon/\epsilon_F$  characterizing spin-polarisation degree vs
 the Bruekner parameter $r_s$ for two typical
 values of screening length $\lambda$: $\lambda/a_0=12.5$(dotted line) and $\lambda/a_0=50$  (solid line).
}
\end{figure}

On obtaining $\delta$ one can easily evaluate the spontaneous magnetic moment per unit area from the relation

\begin{equation}
M={{\mu_B\over{a_0^2}}{\delta(r_s)\over{r_s^2}}}
\end{equation} 

where $\mu_B$ is the Bohr magneton.

To estimate the relative decrease of density of states for electrons adjacent to the Fermi level 
(in a layer ${\hbar}|k_F-k|\le$.${\hbar}/\lambda\,\,\,$ we have considered an expression for exchange energy 
as a function of electron momentum. According to the model we have obtained:

\begin{equation}
\!\!\!\!\epsilon_{ex}(k)\approx{e^2\over{\kappa L}}V_{ex}\!\left({L\over{2\pi}}\right)^2\!\!\!\!\!\int\limits_
{k_F-1/\lambda}^{k_F}\!\!\!\!\!k_1dk_1\!\!\int\limits_0^{2\pi}\theta\left(1/\lambda-\sqrt{k_1^2+k^2-2k_1k 
cos(\alpha)}\right)\!d\alpha
\end{equation} 

The relation (9) originates from a dependence of an electron exchange energy upon spin configuration 
of near-by electrons in k-space.The total energy equals $\epsilon_{tot}(k)+{\hbar}^2k^2/2m+\epsilon_c$
 (the kinetic energy and spin-independent Coulomb energy are added here).
 
From exp. (9) the relative decrease of electron density of states $\rho$ in this energy interval 
 under the Fermi level was deduced as 
 
\begin{equation}
\Delta\rho/\rho\approx 0.12r_s \delta
\end{equation} 

This decrease of density of states (pseudo-gap) is substantial for transport phenomena in 2DEG.

In conclusion, the criterion of a 2DEG to transit to a spin-polarized (magnetized) state was 
numerically derived. The exchange interaction in a many-electron system is modeled by the 
Hartree-Fock description of a pair electron interaction. Spontaneous magnetic moment and a
 decrease of density of states (pseudo-gap) at the Fermi level were also evaluated. The effect
 essentially depends on screening and disorder (dephasing).\\ 

{\bf Acknowledgments}\\
The work was supported through the scientific program "Physics of solid state nanostructures" 
(grant FTN-21(00)-P)  and also through the Russian Basic Research Foundation (grant N 000100397).

\end{document}